# Effects of mesoporous SiO$_2$-CaO nanospheres on the murine peritoneal macrophages / *Candida albicans* interface

R. Diez-Orejas[a*], L. Casarrubios[b], M.J. Feito[b], J.M. Rojo[c], M. Vallet-Regí[d,e], D. Arcos[d,e], M.T. Portolés[b*]

[a] *Departamento de Microbiología y Parasitología, Facultad de Farmacia, Universidad Complutense de Madrid, 28040 Madrid, Spain*

[b] *Departamento de Bioquímica y Biología Molecular, Facultad de Ciencias Químicas, Universidad Complutense de Madrid, Instituto de Investigación Sanitaria del Hospital Clínico San Carlos (IdISSC), 28040 Madrid, Spain*

[c] *Departamento de Medicina Celular y Molecular, Centro de Investigaciones Biológicas, CSIC, 28040 Madrid, Spain*

[d] *Departamento de Química en Ciencias Farmacéuticas, Facultad de Farmacia, Universidad Complutense de Madrid, Instituto de Investigación Sanitaria Hospital 12 de Octubre i+12, Plaza Ramón y Cajal s/n, 28040 Madrid, Spain*

[e] *CIBER de Bioingeniería, Biomateriales y Nanomedicina, CIBER-BBN, Madrid, Spain*

\* Corresponding authors

*E-mail address:* portoles@quim.ucm.es, rosaliad@farm.ucm.es

**Abstract**

The use of nanoparticles for intracellular drug delivery could reduce the toxicity and side effects of the drug but, the uptake of these nanocarriers could induce adverse effects on cells and tissues after their incorporation. Macrophages play a central role in host defense and are responsible for *in vivo* nanoparticle trafficking. Assessment of their defense capacity against pathogenic micro-organisms after nanoparticle uptake, is necessary to prevent infections associated with nanoparticle therapies. In this study, the effects of hollow mesoporous $SiO_2$-CaO nanospheres labeled with fluorescein isothiocyanate (FITC-NanoMBGs) on the function of peritoneal macrophages was assessed by measuring their ability to phagocytize *Candida albicans* expressing a red fluorescent protein. Two macrophage/fungus ratios (MOI 1 and MOI 5) were used and two experimental strategies were carried out: a) pretreatment of macrophages with FITC-NanoMBGs and subsequent fungal infection; b) competition assays after simultaneous addition of fungus and nanospheres. Macrophage pro-inflammatory phenotype markers (CD80 expression and interleukin 6 secretion) were also evaluated. Significant decreases of CD80+ macrophage percentage and interleukin 6 secretion were observed after 30 min, indicating that the simultaneous incorporation of NanoMBG and fungus favors the macrophage non-inflammatory phenotype. The present study evidences that the uptake of these nanospheres in all the studied conditions does not alter the macrophage function. Moreover, intracellular FITC-NanoMBGs induce a transitory increase of the fungal phagocytosis by macrophages at MOI 1 and after a short time of interaction. In the competition assays, as the intracellular fungus quantity increased, the intracellular FITC-NanoMBG content decreased in a MOI- and time-dependent manner. These results have confirmed that macrophages clearly distinguish between inert material and the live yeast in a dynamic intracellular incorporation. Furthermore, macrophage phagocytosis is a critical determinant to know their functional state and a valuable parameter to study the nanomaterial / macrophages */ Candida albicans* interface.

# 1. Introduction

Nanoparticles are currently designed for intracellular drug delivery in order to reduce the toxicity and side effects of the loaded drug to the patient. Silica-based mesoporous nanoparticles present high loading capacity within their pores and the possibility of modifying their surface to target the damaged tissues [1]. However, depending on the characteristics of these nanocarriers, their interaction with tissues and cells could induce adverse effects [2]. Thus, recent studies evidence that micro- and nanoparticles can produce immunosuppression that initially limits early inflammation but contributes later to fibrosis, cancer, and infection [3]. Macrophages are among the first cell types which take up nanoparticles [4-8] and they are primarily responsible for their trafficking *in vivo* [9,10]. These phagocytic cells play a central role in host defense, removal of dead cells and pathogens, inflammatory response regulation and modulation of adaptive immunity [11]. Macrophages have functional plasticity that allows them to respond to their local environment, changing their phenotype between two extremes defined as pro-inflammatory (M1) and anti-inflammatory (M2) phenotypes, characterized by different cell surface markers and cytokines [12]. Thus, the evaluation of the *in vitro* response of macrophages after exposure to nanomaterials is an important issue due to the key role of these phenotypes in both the host response and the inflammation/healing processes [13].

Phagocytosis by macrophages is a critical process for the uptake and degradation of infectious agents and senescent cells, and it participates in tissue remodeling, immune response, and inflammation [14]. A defective phagocytic ability of macrophages is considered as a critical determinant in the progression of different diseases because this fact is related to a suppressive adaptive immune response [15,16].

On the other hand, the evaluation of the possible macrophage capability impairment against pathogens after nanoparticle uptake is necessary to prevent infections that could occur during treatments based on the administration of these nanocarriers. In this sense, different strategies have been developed to minimize risk of infections related with the use of biomedical materials and their complications [17].

*Candida albicans* is a fungal pathogen that easily colonizes host tissues and implant devices forming mature biofilms directly involved in pathogenesis and drug resistance [18]. On the other hand, the pathogenicity of this opportunistic human fungal pathogen depends on the competence of host immune response that if diminishes, *C. albicans* behaves like a true pathogen. Both innate and adaptive components are involved in the immune response to microbial pathogens [19]. Thus, although the innate immune response plays a key role in the initial containment of the fungus, both cellular and humoral responses are needed to finally finish the infection [20-24]. The importance of phagocytes in the immune response against the fungus and the resolution of the infection has been accurately demonstrated [25].

Phagocytosis is a very important step by which macrophages destroy pathogens and a huge amount of scientific data defines this process as a very complex and essential phase for the immune response [26,27]. In this context, it is relevant to note that many pathogens have evolved strategies to avoid or subvert phagocytosis at various stages of this process [28,29]. Similarly, *C. albicans* has developed different mechanisms and one of these well-known strategies is the morphogenetic switch inside macrophages [30,31]. Other important fungal evasion strategies have been described [32,33], indicating the necessity to improve antifungal treatments against this pathogen.

In the present study, we have analyzed by flow cytometry and confocal microscopy the phagocytic capacity of murine peritoneal macrophages after treatment with mesoporous $SiO_2$-CaO nanospheres labeled with fluorescein isothiocyanate (FITC-NanoMBGs) and subsequent infection with the fungus *Candida albicans* expressing a red fluorescent protein (RFP-*Candida albicans*) at two macrophage/fungus ratios (multiplicity of infection, MOI): MOI 1 and MOI 5. We have also carried out competition assays, after simultaneous addition of fungus and nanoparticles to peritoneal macrophages, in order to study the modulation of intracellular FITC-NanoMBG content by intracellular RFP-*Candida albicans* yeasts or *vice versa*. In these assays, we have also analyzed the effects of FITC-NanoMBGs and *Candida albicans* on macrophage polarization towards the pro-inflammatory M1 phenotype by measuring the CD80 expression (as specific M1 marker) and the secretion of interleukin-6 (IL-6, as pro-inflammatory cytokine).

It is important to highlight that the biomedical treatments with nanoparticles require the study of the effects of nanomaterials on the macrophage / microbial pathogen interface to know if the intracellular uptake of these nanoparticles could alter the immune response. On the other hand, the present study will allow us to obtain information for the possible future biomedical employment of these mesoporous $SiO_2$-CaO nanospheres in an infection scenario. We have recently shown the utility of this nanomaterial for intracellular drug delivery [34,35]. These previous works have opened the way for future potential profits of this nanomaterial for antifungal delivery by the analysis of the uptake by macrophages of both nanospheres and *C. albicans*. The present work will contribute for the better understanding of the functional state of macrophages after treatment with these potential nanocarriers.

## 2. Materials and methods

*2.1. Synthesis and characterization of FITC-NanoMBGs*

The synthesis of hollow mesoporous $SiO_2$-CaO nanospheres (NanoMBGs) was carried out by the method described by Li et al. [36], using a double template strategy to obtain particles with a hollow core surrounded by a shell containing a radial porous structure (NanoMBGs). For the subsequent internalization studies, NanoMBGs were labelled with FITC thus obtaining FTIC-NanoMBGs particles (see supporting information for a complete description of the synthesis and labelling methods).

NanoMBGs were characterized by scanning electron microscopy (SEM) using a JEOL F-6335 microscope (JEOL Ltd., Tokyo, Japan), operating at 20 kV and equipped with an energy dispersive X-ray spectrometer (EDS). Previously, the samples were mounted on stubs and gold coated in vacuum using a sputter coater (Balzers SCD 004, Wiesbaden- Nordenstadt, Germany). Transmission electron microscopy (TEM) was carried out using a JEOL-1400 microscope, operating at 300 kV (Cs 0.6mm, resolution 1.7 Å). Images were recorded using a CCD camera (model Keen view, SIS analyses size 1024 X 1024, pixel size 23.5mm x 23.5mm) at 60000X magnification using a low-dose condition. Nitrogen adsorption/desorption isotherms were obtained with an ASAP 2020 porosimeter. NanoMBGs were previously degassed under vacuum for 15 h, at 150 ºC. The surface area was determined using the Brunauer-Emmett-Teller (BET) method. The pore size distribution between 0.5 and 40 nm was determined from the adsorption branch of the isotherm by means of the Barret-Joyner-Halenda (BJH) method. The surface area was calculated by the BET method and the pore size distribution was determined by the BJH method using the adsorption branch of the isotherm. Particle size was determined from SEM micrographs using the intercept technique.

*2.2. Isolation and culture of murine peritoneal macrophages*

Primary macrophages were obtained from the peritoneum of untreated mice as described elsewhere [37]. Briefly, naive non treated CL57/BL mice were killed, and the skin was removed from the abdominal area. Mice were then injected intraperitoneally with 4-5 mL of phosphate buffered saline (PBS) using an 18 gauge needle. Without extracting the needle, the abdomen was gently massaged and then as much fluid from the peritoneum as possible was slowly withdrawn with the syringe. After removing, the peritoneal cells were gently washed with PBS before use. All procedures were approved by Institutional Animal Care and Use Committees. Mouse peritoneal macrophages ($10^5$ cells/mL) were seeded in 6 well culture plates (CULTEK S.L.U., Madrid, Spain) in complete medium. This medium contains Dulbecco's Modified Eagle Medium (DMEM) supplemented with 10% fetal bovine serum (FBS, Gibco, BRL), 1 mM L-glutamine (BioWhittaker Europe, Belgium), penicillin (200 μg/mL, BioWhittaker Europe, Belgium) and streptomycin (200 μg/mL, BioWhittaker Europe, Belgium) at 37 ºC under a 5% $CO_2$ atmosphere.

*2.3. Candida albicans strains*

The *C. albicans* strains used in this study were the clinical isolate SC5314 and the CAF2-dTOM2 derived from SC5314 expressing a red fluorescent protein (RFP) called RFP-*Candida albicans* [38], kindly provided by Dr. D. Prieto and Dr. J. Pla. This red fluorescent strain was synthesized by codon optimization of the DsRed-derived RFP dTomato gene [39] using the tetracycline-dependent integrative plasmid pNIM1R. Expression in *C. albicans* was also detectable in SD plates as pink-reddish colored colonies after two days of growth [40]. Yeast strain was short-term stored at 4°C and grown at 37°C in YPD medium (2% glucose, 2% peptone, 1% yeast extract) plus amino acids and chloramphenicol (10 μg/mL) for 48 hours.

*2.4. Treatment of murine peritoneal macrophages with FITC-NanoMBGs for 24 h and subsequent infection with RFP-Candida albicans for different times and multiplicity of infection*

Prior to *Candida* infection, murine peritoneal macrophages were seeded a) onto 24 well culture plates (for confocal microscopy studies), containing sterile 12-mm-diameter round glass coverslips, at a density of $5 \times 10^5$ cells per well, and b) onto 6 well culture plates (for flow cytometry studies), at a density of $1 \times 10^6$ cells per well. Then, macrophages were treated with a FITC-NanoMBGs suspension (50 µg/mL), which involves approximately $5.3 \cdot 10^9$ FITC-NanoMBGs/mL, for 24 h at 37 ºC under a 5% $CO_2$ atmosphere. After treatment, some wells were washed to retire the nanomaterial, meanwhile in others the nanomaterial was not removed, in order to compare the effects produced by the previous uptake of nanospheres and by the continuous presence of them. Controls with macrophages without FITC-NanoMBG treatment were carried out in parallel. On the other hand, RFP- *C. albicans* was grown overnight at 30 ºC on solid YED medium (to maintain cells in the yeast form). Yeast cells, harvested and washed twice with phosphate buffered saline (PBS), were counted and diluted to the desired density in complete culture media.

In all the conditions described previously, macrophages were infected for different times (30, 45 and 90 min) with *C. albicans* using two macrophage/fungus ratios (multiplicity of infection, MOI): MOI 1 and MOI 5. All the experiments were performed at 37 °C and 5% $CO_2$. We employed controls with macrophages without *C. albicans* infection or without nanoesphere treatment in parallel.

*2.5. Infection of murine peritoneal macrophages with RFP-Candida albicans and simultaneous treatment with FITC-NanoMBGs: competition assay*

Similarly to the previous experimental conditions, murine peritoneal macrophages were seeded either onto 24 well culture plates (for confocal microscopy studies) and onto 6 well culture plates (for flow cytometry studies). Then, macrophages were simultaneously treated with both the fungus RFP-*Candida albicans* (MOI 1 and MOI 5) and the nanomaterial (FITC-NanoMBGs, 50μg/mL) for different times (30, 45 and 90 min) at 37 ºC under a 5% $CO_2$ atmosphere. We designed this experiment, considered as a competition assay, to evaluate the possible modulation of both macrophage fungal phagocytosis and FITC-NanoMBG uptake by the simultaneous presence of the nanomaterial and the yeasts, respectively. We employed controls with macrophages without *C. albicans* infection or without nanoesphere treatment in parallel.

*2.6. Uptake of FITC-NanoMBGs and phagocytosis of RFP-Candida albicans by murine peritoneal macrophages evaluated by flow cytometry and confocal microscopy*

To evaluate FITC-NanoMBG uptake and RFP-*Candida albicans* phagocytosis by murine peritoneal macrophages by flow cytometry, cells were washed three times with ice-cold phosphate buffered saline (PBS) and harvested using cell scrapers. Then, cell suspensions were centrifuged at 300xg for 5 min and the pellets were suspended in 300 μl of PBS with 2% FBS. The fluorescence of FITC-NanoMBGs and RFP-expressing *C. albicans* was excited at 488 nm and measured with 530/30 and 585/42 band pass filters, respectively in a FACScalibur Becton Dickinson flow cytometer. The conditions for the data acquisition and flow cytometric analysis were established using negative and positive controls with the CellQuest Program of Becton Dickinson and these conditions were maintained during all the experiments. Each experiment was carried out three times and single representative experiments are displayed. For statistical significance, at least $10^4$ cells were analyzed in each sample.

A flow cytometry diagram is shown below as an example of the analysis carried out to quantify the uptake of FITC-NanoMBGs and phagocytosis of RFP-*Candida albicans* by murine peritoneal macrophages.

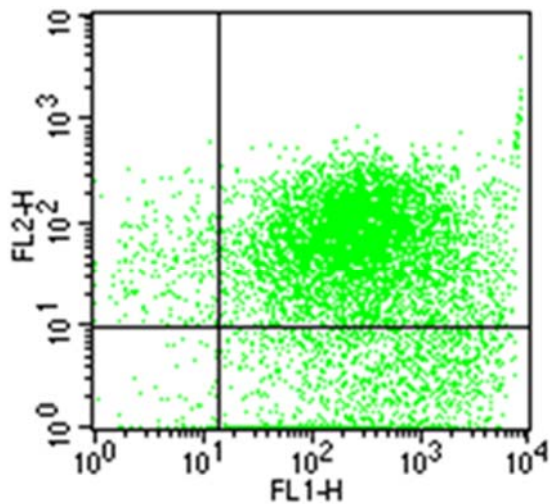

**Example of the flow cytometric analysis of FITC-NanoMBG uptake and RFP-*Candida albicans* phagocytosis by murine peritoneal macrophages.** FL1 (green fluorescence) corresponds to the fluorescence of FITC-NanoMBGs inside the macrophages. FL2 (red fluorescence) corresponds to the fluorescence of RFP-*Candida albicans* inside the macrophages. The vertical line corresponds to the limit value of basal fluorescence FL1 (cells without FITC-NanoMBGs). The horizontal line corresponds to the limit value of basal fluorescence FL2 (cells without RFP-*Candida albicans*). Thus, the four quadrants in the graph correspond to the four detected populations of macrophages:
UL (Up Left) = macrophages without intracellular FITC-NanoMBGs and with RFP-*Candida*;
UR (Up Right) = macrophages with intracellular FITC-NanoMBGs and with RFP-*Candida;*
LL (Low Left) = macrophages without intracellular FITC-NanoMBGs and without RFP-*Candida;* LR (Low Right) = macrophages with intracellular FITC-NanoMBGs and without RFP-*Candida.* In the table, the percentage of phagocytosis corresponds to the % of macrophages that presented the fungus inside and therefore exhibited red fluorescence above the horizontal line (UL% + UR%). The percentage of FITC-NanoMBG uptake corresponds to the % of macrophages that presented nanospheres inside and therefore exhibited green fluorescence to the right of the vertical line (UR% + LR%). X Mean and Y Mean correspond to the green and red fluorescence values respectively, in arbitrary units, emitted by each cell population (each quadrant).

For confocal microscopy studies, cells were fixed with 3.7% paraformaldehyde in PBS and observed using a Leica SP2 Confocal Laser Scanning Microscope. The fluorescence of RFP-expressing *C. albicans* was excited at 488 nm and measured at 584/663 nm. DAPI fluorescence was excited at 405 nm and measured at 409/468 nm. FITC fluorescence of FITC-NanoMBGs was excited at 488 nm and measured at 505/547 nm.

*2.7. Effects of FITC-NanoMBGs and Candida albicans on macrophage polarization towards M1 phenotype*

In order to assess the effects of FITC-NanoMBGs and *Candida albicans* on macrophage polarization towards M1 phenotype, the CD80 expression as specific marker of pro-inflammatory M1 macrophages [8,41] was evaluated after two experimental approaches.

a) 24 hours pretreatment of macrophages with FITC-NanoMBGs, elimination of the nanomaterial by washing and then, incubation for 30 min with *Candida albicans* at the desired density (MOI 1 or MOI 5);

b) Simultaneous treatment of macrophages with FITC-NanoMBGs and *C. albicans* (MOI 1 or MOI 5) during 30 min.

In parallel, we employed control cells in the absence of nanomaterial or/and fungus. For flow cytometric CD80 detection, after detachment and centrifugation, cells were incubated in 45 µl of staining buffer (PBS, 2.5% FBS Gibco, BRL and 0.1% sodium azide, Sigma-Aldrich Corporation, St. Louis, MO, USA) with 5 µl of normal mouse serum inactivated for 15 min at 4º C in order to block the Fc receptors on the macrophage plasma membrane, before adding the primary antibody, and to prevent non-specific binding. Then, cells were incubated with phycoerythrin (PE) conjugated anti-mouse CD80 antibody (2.5 µg/mL, BioLegend, San Diego, California) for 30 min in the dark. Then, labelled cells were analyzed using a FACSCalibur Becton Dickinson flow cytometer. PE fluorescence was excited at 488 nm and measured at 585/42 nm. The conditions for data acquisition and analysis were established using negative and positive controls with the CellQuest Program of Becton Dickinson, and these conditions were maintained in all the experiments. Each experiment was carried out three times and single representative experiments are displayed. For statistical significance, at least $10^4$ cells were analyzed in each sample.

The amount of IL-6 in the culture medium was quantified by ELISA (Gen-Probe, Diaclone) with pre-coated strip plates, biotinylated secondary antibodies and streptavidin-avidin conjugated to horseradish peroxidase for colorimetric reaction, according to the manufacturer's instructions. Recombinant cytokine was used as standard. The absorbance at 450 nm of the samples and standards was measured using an ELISA Plate Reader. The sensitivity of this assay was 10 pg/mL and its inter-assay variation coefficient was < 10%.

*2.8. Statistics*

Data are expressed as means ± standard deviations of a representative of three experiments carried out in triplicate. Statistical analysis was performed using the Statistical Package for the Social Sciences (SPSS) version 19 software. Statistical comparisons were made by analysis of variance (ANOVA). Scheffé test was used for *post hoc* evaluations of differences among groups. In all of the statistical evaluations, $p < 0.05$ was considered as statistically significant.

# 3. Results and discussion

*3.1. Physical-chemical characterization of FITC-NanoMBGs*

SEM and TEM images (Figure 1) show that NanoMBG material consists on monodisperse spherical nanoparticles of around 250 nm in diameter, although some nanoparticles have certain degree of polyhedral morphology and exhibiting hollow core- radial shell porous structure. EDX spectroscopy carried out during TEM observation revealed a chemical composition of 81.5 $SiO_2$- 18.5 CaO (% mol) for NanoMBG.

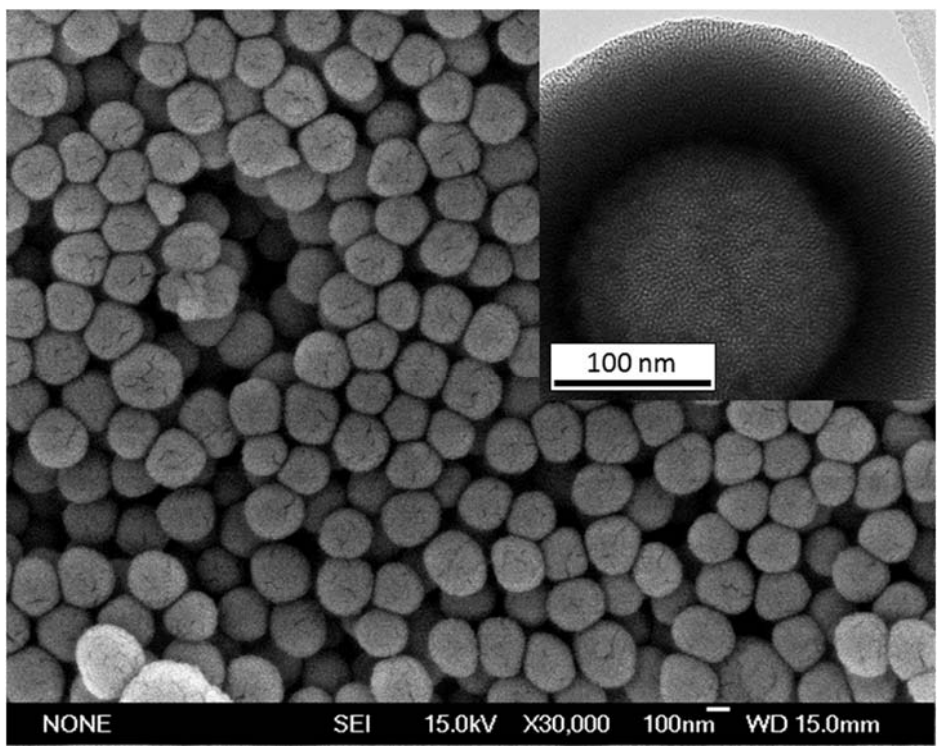

**Figure 1. SEM micrograph on NanoMBG spheres.** The inset shows a TEM image evidencing the hollo core-radial shell porous structure of the spheres. Nitrogen adsorption analysis confirm the highly porous structure of NanoMBG spheres, showing high values of surface area (more than 540 $m^2·g^{-1}$) and pore volume (see Table 1).

**Table 1. Surface area, pore volume, pore size and material density obtained by nitrogen adsorption analysis for NanoMBG.**

|  | Surface area $(m^2·g^{-1})$ | Porosity $(cm^3g^{-1})$ | Pore size (nm) | Density $(g·cm^{-3})$ |
|---|---|---|---|---|
| **NanoMBG** | 543.6 | 0.435 | 2.2 | 1.186 |

*3.2. Influence of previous 24 h FITC-NanoMBG treatment on RFP-Candida albicans phagocytosis by murine peritoneal macrophages*

In order to know if the uptake or/and the presence of FITC-NanoMBGs in the culture medium alter the function of murine peritoneal macrophages, we analyze RFP-*Candida albicans* phagocytosis by these cells after 24 h of treatment with this nanomaterial by performing two experimental approaches. Thus, after FITC-NanoMBG treatment, some wells were washed to eliminate the nanomaterial and fresh medium was added, meanwhile other wells were maintained with nanospheres in the medium. After these two treatments, a fungal phagocytosis assay using a RFP-*Candida albicans* strain was settled down.

Phagocytosis of *Candida* by macrophages depends on both the ratio between the fungus and the macrophages (defined as multiplicity of infection or MOI) and the time of interaction. Thus, fungal phagocytosis increases in a MOI- and time-dependent manner. In this sense, it is generally accepted that 90-120 min of macrophage-fungus interaction is the final time course for fungal phagocytosis (at this time point macrophages are not be able to phagocyte more fungus). Therefore, the conditions we have employed for these phagocytosis studies have been settled down taking into account previous studies [42-45].

In the present work we have analyzed fungal phagocytosis by peritoneal macrophages at two different MOIs (1 and 5) and after 30, 45 and 90 min of infection. We detected by flow cytometry the percentage of macrophages with red fluorescence as a measure of the phagocytosis percentage. To quantify only the yeasts inside macrophages, we employed Trypan blue, incapable of penetrating into viable phagocytes, to quench the fluorescence of the outer RFP-labeled yeasts [46]. We compare the fungal phagocytosis in these conditions with control macrophages (not treated with nanospheres).

Figure 2 shows different percentages of phagocytosis depending on the MOI and the interaction time. After FITC-NanoMBG uptake and at the shortest time of interaction with the fungus (30 min) at MOI 1 (Figure 2A), we detected significant increases of the phagocytic capability of macrophages (percentage of phagocytosis), depending on the presence of the nanomaterial in the medium (33%, non-washed spheres) or its no presence (42%, washed spheres) in comparison with control macrophages (25%). At MOI 1 and after 45 min of interaction, the treatment with nanospheres diminished the phagocytosis capability of macrophages significantly from 60% (control) to 51% (non-washed and washed spheres). Finally, when fungal phagocytosis was analyzed after 90 min of interaction, the previous treatment with nanospheres decreased the phagocytosis from 70% (control) to 66% (non-washed spheres) and 59% (washed spheres), but this effect was not statistically significant (Figure 2A).

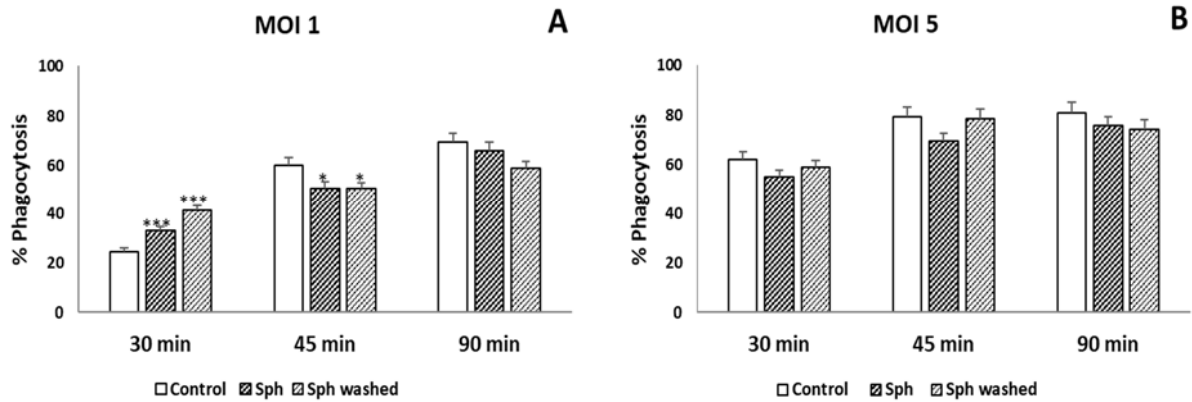

**Figure 2. Flow cytometric analysis of RFP-*Candida albicans* phagocytosis by peritoneal macrophages after previous treatment of 24 h with FITC-NanoMBGs (50μg/mL).** The percentage of phagocytosis corresponds to the % of macrophages that presented the fungus inside and therefore exhibited red fluorescence. Fungal infection was assayed for different times of interaction (30, 45 and 40 min) with MOI 1 (Figure 2A) and MOI 5 (Figure 2B). The different barred bars indicate the presence (Sph) or the absence (Sph washed) of FITC-NanoMBGs in the culture medium. White bars correspond to control macrophages (not treated with nanospheres). Statistical significance: ***$p<0.005$, *$p<0.05$ (comparison between control and FITC-NanoMBG treated macrophages at the same MOI and time of interaction).

Figure 2B shows the percentage of fungal phagocytosis with MOI 5 after 30, 45 and 90 min of interaction. At 30 and 45 min of infection and at this high infection ratio, the presence of nanospheres in the medium decreased the percentage of phagocytosis, although this effect was not statistically significant. At 90 min of infection, there was no differences in fungal phagocytosis independently of the presence or the absence of the nanomaterial in the medium, probably due to the high number of yeasts.

The phagocytosis capability of control macrophages increases along the time and depending on the number of yeasts present in the medium (MOI employed), as it is shown in Figures 2A and 2B. The results obtained after FITC-NanoMBG treatment allow us to affirm that these nanospheres do not exhibit toxicity on the macrophage function in the tested conditions because their phagocytosis capability was time- and MOI-dependent after treatment with these nanospheres, in agreement with previous studies [6]. Interestingly, this macrophage function increases after FITC-NanoMBG treatment at the shortest time of fungal interaction and MOI 1, thus suggesting an initial stimulation induced by these nanospheres on the activity of peritoneal macrophages that decreases along the time, probably due to the nanomaterial uptake (Figure 2A). Moreover, in these experiments with and without the presence of nanospheres, we did not detect significant differences in fungal phagocytosis, thus indicating that macrophages discern between the nanomaterial and the fungus.

*3.3. RFP-Candida albicans phagocytosis and FITC-NanoMBG uptake by macrophages in a competition assay*

We performed a competition assay in order to evaluate the possible modulation of both RFP-*Candida albicans* phagocytosis and FITC-NanoMBG uptake by macrophages due to the simultaneous presence of the nanomaterial and the yeasts, respectively. In this kind of assay, we cultured murine peritoneal macrophages with FITC-NanoMBGs (50μg/mL) and fungal cells (at the desired MOI) simultaneously during different times (30 min, 45 min and 90 min) and we detected intracellular RFP-*Candida albicans* and FITC-NanoMBGs by flow cytometry and confocal microscopy as described above. We carried out controls with macrophages without *C. albicans* infection or without nanomaterial treatment in parallel.

As it is shown in Figure 3A, after 30 min of the competition assay and at MOI 1, we detected a significant increase of the phagocytic capability of macrophages (45%) induced by the presence of nanospheres comparing with the of control macrophages without nanomaterial (25%). However, after 45 min and 90 min of interaction at MOI 1, fungal phagocytosis by macrophages diminished in the presence of nanospheres from 60% and 70% phagocytosis by control macrophages to 50% and 61% phagocytosis, respectively (Figures 3B and 3C). When employing a higher macrophage/fungus ratio (MOI 5), the presence of nanospheres also induced a significant fungal phagocytosis increase at the shortest time of interaction (30 min, Figure 3A). Thus, phagocytosis by control macrophages at MOI 5 was around 63% (white bar) and in the competition assay there was a phagocytosis stimulation to 72% (barred bar).

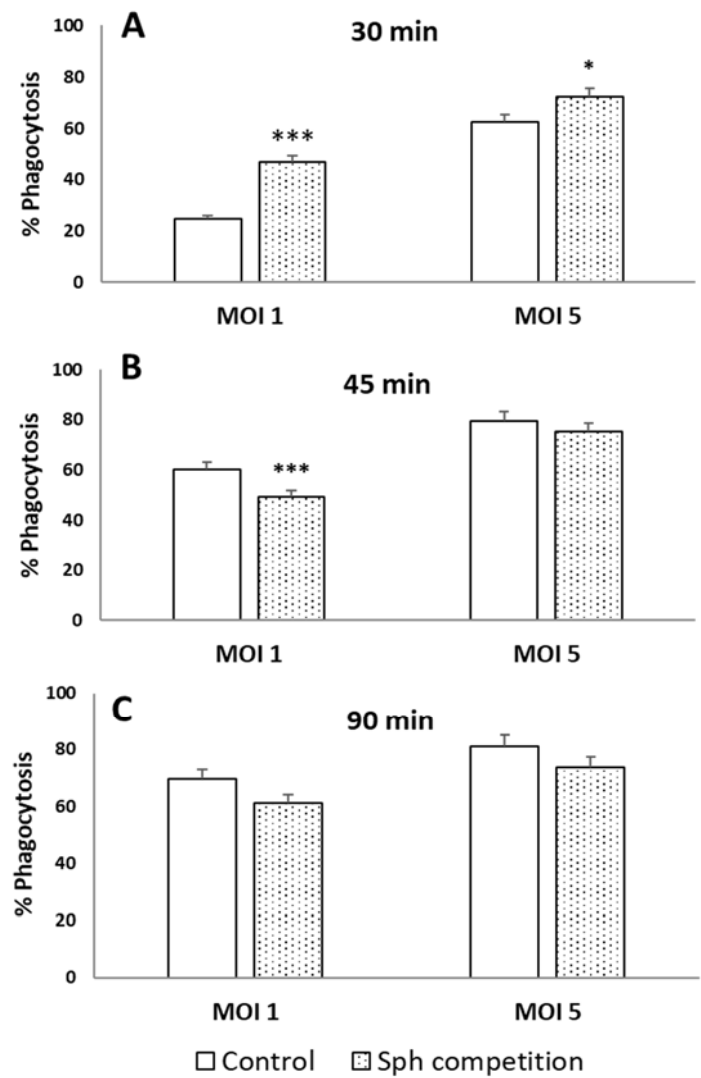

**Figure 3. Flow cytometric analysis of RFP-*Candida albicans* phagocytosis by peritoneal macrophages in the presence of FITC-NanoMBGs (50μg/mL).** The barred bars indicate the percentage of fungal phagocytosis at different MOIs (1 and 5) during the competition assay in the presence of nanospheres for different times (30, 45 and 90 min). White bars correspond to control macrophages (not treated with nanospheres). Statistical significance: ***$p<0.005$ (comparison between macrophages with or without nanosphere treatment at the same MOI and time of interaction).

At longer times the phagocytosis of the fungus in this competition assay diminished due to the presence of the nanospheres, obtaining values from 79% in control to 75% with nanospheres at 45 min, and from 81% in control to 74 % with nanospheres at 90 min (Figures 3B and 3C). This decrease observed at 45 min and 90 min, is in agreement with the effects described in section 3.2, using previously treated macrophages with nanospheres before RFP-*Candida albicans* infection (Figure 2). However, the simultaneous presence of both nanomaterial and fungus induced an initial stimulation at 30 min on the phagocytosis macrophage capability that always occurs independently of the MOI employed.

In a deeper analysis, we compared the distribution of the three different macrophage populations along the time (30, 45 and 90 min) after the simultaneous addition of FITC-NanoMBGs and RFP-*C. albicans* at two macrophage/fungus ratios (MOI 1 and MOI 5):

- Macrophages with intracellular FITC-NanoMBGs and without RFP-*Candida* (+Sph/-Cand)

- Macrophages without intracellular FITC-NanoMBGs and with RFP-*Candida* (-Sph/+Cand)

- Macrophages with intracellular FITC-NanoMBGs and with RFP-*Candida* (+Sph/+Cand).

Figures 4A and 4B show that only a few macrophages without nanospheres incorporated the fungus along the time (-Sph/+Cand population) and independently of the MOI employed. Thus, Figure 4A (MOI 1) shows a percentage of -Sph/+Cand population of 0.3%, 0.17% and 0.45% and in Figure 4B (MOI 5) percentages of 2%, 3 % and 4.7% were observed corresponding to this population at the different times of treatment. On the other hand, when we compared +Sph/+Cand and +Sph/-Cand macrophage populations, we detected a different behavior along the interaction time and related to the MOI employed. With respect to the results obtained at MOI 1 (Figure 4A), around 98% of the macrophages incorporated NanoMBGs, thus reflecting the high uptake of this nanomaterial and its maintenance along the time. The +Sph/+Cand population (cells that incorporated the nanomaterial and phagocytosed the fungus) increased significantly along the time (47%, 50% and 60%). These results point out that the uptake of nanospheres by macrophages does not avoid fungal phagocytosis, and *vice versa*. Moreover, the continuous presence of both stimuli favors the fungal phagocytosis and as a consequence, the +Sph/-Cand population (cells that has incorporated the nanomaterial but not the fungus) decreased significantly along the time (52%, 49% and 36%). Thus, at the longest time of interaction of 90 min, we can observe a significant increase ($p < 0.005$) of the +Sph/+Cand macrophages, revealing that the macrophage capability of phagocytosis is not altered by the uptake of these nanospheres.

With respect to the results obtained at MOI 5 (Figure 4B), when a higher dose of *C. albicans* is employed but using the same quantity of the nanomaterial, we detected higher percentage of the +Sph/+Cand population (comparing to MOI 1) that is maintained along the time (70%, 72% and 69%) and lower values of the +Sph/-Cand population (27%, 24% and 22%).

To conclude, the results represented in Figure 4 point out that the presence of nanomaterial does not alter the phagocytic capacity of macrophages. Moreover, the nanomaterial has a stimulating capability, reflected not only by the increase of the macrophage phagocytosis along the time but also directly related to the quantity of the fungus present in the medium. In this sense, we want to highlight that the macrophage population with only the fungus inside is practically undetectable, even at MOI 5. These results point out the possible benefits of these nanospheres for antifungal delivery.

To complete these competition assays, we also observed by confocal microscopy both the fungal phagocytosis and nanomaterial incorporation after the simultaneous addition of FITC-NanoMBGs and RFP-*C. albicans* with MOI 1 and MOI 5 (Figures 5 A y B). This technique allows detailed viewing on individual macrophages infected with the fungus in red and the intracellular nanomaterial in green. Magnification of these images shows differences depending on the number of yeasts in the medium. Thus, at MOI 1 (Figure 5 A), we can observe both the yeast and the filaments of *Candida albicans* inside macrophages and at MOI 5 (Figure 5B) macrophages exhibit a high number of phagocytosed fungal cells in comparison with MOI 1. These images are in agreement with the results described in Figures 4A and 4B.

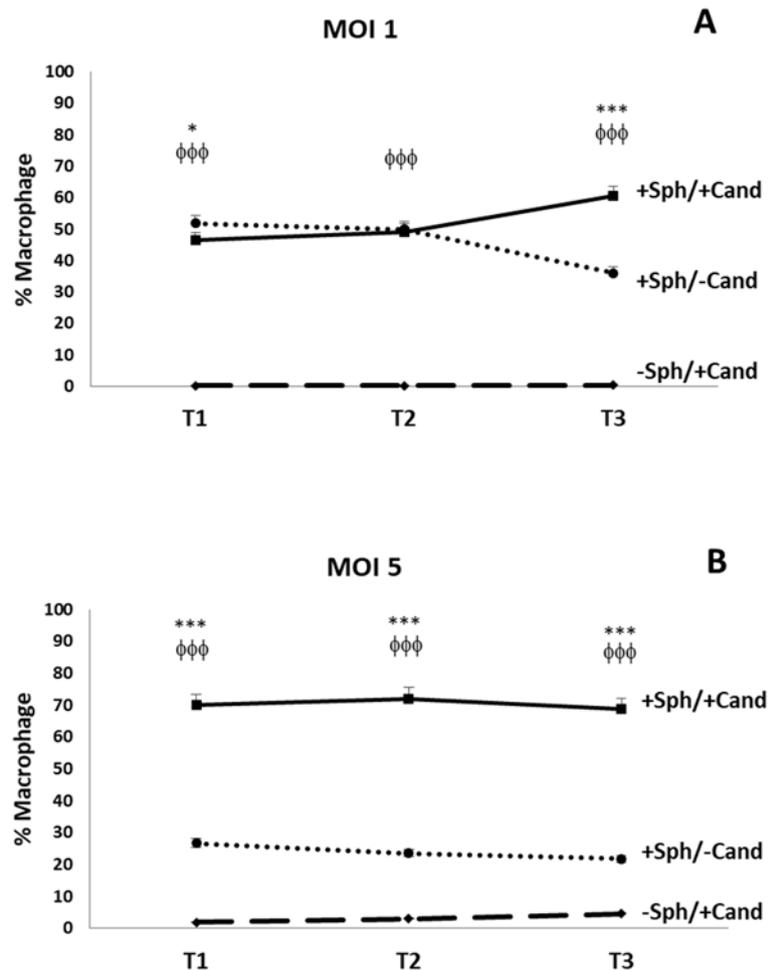

Figure 4. Flow cytometric analysis of three different macrophage populations at 30 min (T1), 45 min (T2) and 90 min (T3) after the simultaneous addition of FITC-NanoMBGs (50μg/mL) and RFP-*Candida albicans* at MOI 1 (A) and MOI 5 (B). Evolution along the time of three macrophage populations represented by the percentage of cells with intracellular FITC-NanoMBGs (+Sph) or/and RFP-*Candida albicans* (+Cand): +Sph/-Cand (macrophages with intracellular FITC-NanoMBGs and without RFP-*Candida*), -Sph/+Cand (macrophages without intracellular FITC-NanoMBGs and with RFP-*Candida*) and +Sph/+Cand (macrophages with intracellular FITC-NanoMBGs and with RFP-*Candida*). Statistical significance of different comparisons among populations: ***$p<0.005$, *$p<0.05$ (comparison between +Sph/+Cand and +Sph/-Cand macrophages at the same MOI and time of interaction). ɸɸɸ$p<0.005$ (comparison between +Sph/+Cand and -Sph/+Cand macrophages at the same MOI and time of interaction).

Moreover, in order to quantify the intracellular nanomaterial and the number of yeasts ingested in this competition assay, we analyzed by flow cytometry both green and red fluorescence intensities. Bright field and dark field images were obtained by confocal microscopy in order to better identify the uptake inside the cells. Figures 5 A, 5 B, 5 C and 5 D correspond to bright field and fluorescence images of single macrophages showing intracellular FITC-NanoMBGs (in green), phagocyted RFP-*C. albicans* (in red, A and B at MOI 1, C and D at MOI 5), nucleus (in blue) and extracellular nanospheres in the medium. Figures 5 E and 5 F show the Fl 1 (green) and Fl 2 (red) values that allow us to quantify the intracellular nanospheres (FL1) and fungus (FL2) along the time of interaction (30, 45 and 90 min) at the two MOIs employed. The confocal images clearly show that the fluorescence of FITC-NanoMBGs is maintained during the assays, allowing the identification of the nanospheres inside the cells. Veeranarayanan *et al.* demonstrated the high biocompatibility and photostability of FITC labeled silica nanoparticles incorporated by endothelial cells (HUVEC) up to the fifth day of culture, highlighting the importance of the use of fluorescent nanomaterials in the field of medical imaging as highly efficient cell tags for diagnostic targeting and therapeutic delivery [47]. After the fifth day of culture, a decrease in the fluorescence intensity was observed and was attributed to the high mitotic index of HUVEC cells, that dilutes the nanoparticles per cell and leads to reduced fluorescence intensity. In our study, the maximum time of the assay was 90 min. This time is short enough so that the photoluminescence of the nanospheres is not affected. As these authors indicates, another notable factor for loss of fluorescence can be attributed to the efficient exocytosis of these nanomaterials by HUVEC cells [47].

Regarding the quantity of the nanomaterial (FL1) shown in Figure 5 E, +Sph/-Cand macrophages maintains the initial high intracellular uptake of nanospheres along the time. Meanwhile, in +Sph/+Cand population, a significant decrease in the initial amount of the intracellular nanomaterial is detected at 30 min related to the multiplicity of fungal infection employed. These data indicate that a lower number of nanospheres are incorporated by the cells due to the fungus phagocytosis, and this effect is more pronounced when the MOI increases. At longer times (45 min and 90 min), green fluorescence intensity significantly diminishes evidencing the exocytosis of this nanomaterial is directly associated to the quantity of *C. albicans*. These results (Figure 5 E) point out the necessity of the macrophage to exocytose some nanospheres in order to sustain its fungal phagocytosis capability depending on the number of yeasts in the medium.

The quantity of the fungus ingested by macrophages in these conditions is shown in Figure 5 F as red fluorescence intensity (FL2). Fungal phagocytosis of control macrophages without nanomaterial treatment is shown as -Sph/+Cand population at MOI 1 and MOI 5. At 30 min (T1), the number of the ingested yeasts increases in macrophages with nanospheres when comparing control macrophages. However, at longer times (T2 and T3), the presence of the intracellular nanomaterial diminishes the number of ingested yeasts compared with control macrophages.

These results evidence that NanoMBG internalization by macrophages does not alter their phagocytic capacity because these cells are still able to phagocyte the fungus and, at early time points, there is an increase in their functional capability. Moreover, the presence of extracellular fungus modulates the function of macrophages, pointing out the necessity of these cells to exocytose the nanomaterial in order to increase their fungal uptake. The characteristics of this nanomaterial suggest their potential utility for drug delivery without affecting the innate immune response in the context of a fungal infection.

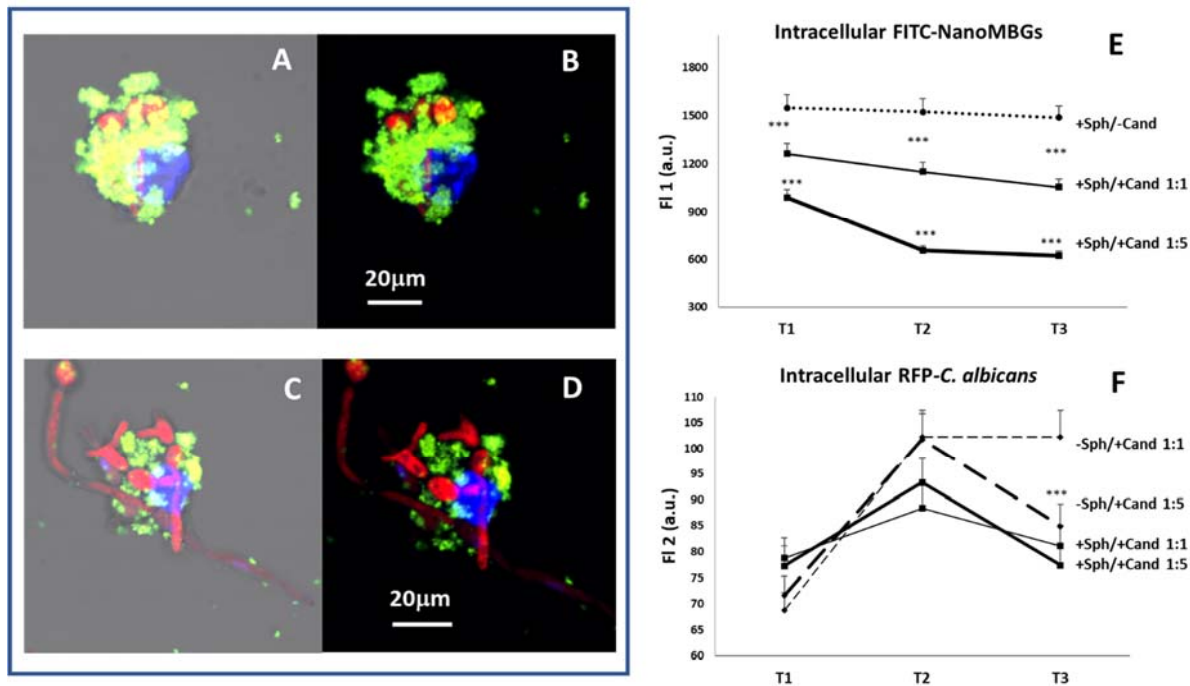

**Figure 5. Confocal microscopy images and flow cytometric analysis of intracellular FITC-NanoMBGs and RFP-*Candida albicans* of peritoneal macrophages after the simultaneous addition of these nanospheres (50µg/mL) and the fungus at MOI 1 (1:1) and MOI 5 (1:5).** Figures A, B, C and D correspond to bright field and fluorescence images of single macrophages showing intracellular FITC-NanoMBGs (in green), phagocyted RFP-*C. albicans* (in red, A and B at MOI 1, C and D at MOI 5), nucleus (in blue) and extracellular nanospheres in the medium. Figure E corresponds to the green fluorescence intensity of intracellular FITC-NanoMBGs (FL1). Figure F corresponds to the red fluorescence intensity of intracellular RFP-*Candida albicans* (FL2). Figures E and F show the FL1 and FL2 of different macrophage populations:
+Sph/-Cand (macrophages with intracellular FITC-NanoMBGs and without RFP-*Candida*)
-Sph/+Cand (macrophages without intracellular FITC-NanoMBGs and with RFP-*Candida*)
+Sph/+Cand (macrophages with intracellular FITC-NanoMBGs and with RFP-*Candida*).
Statistical significance: ***$p<0.005$.

*3.4. Effects of FITC-NanoMBGs and Candida albicans on macrophage polarization towards M1 phenotype*

Given the importance of the balance between the M1-mediated inflammation and M2-mediated regeneration after the treatment with any nanomaterial, we have evaluated the macrophage polarization process after nanospheres treatment. Thus, to assess the effects of FITC-NanoMBGs and *Candida albicans* on macrophage polarization towards M1 phenotype, we detect by flow cytometry the CD80 expression, as specific marker of pro-inflammatory M1 macrophages. In addition, we also quantify by ELISA the interleukin-6 (IL-6) levels in the culture medium. Two approaches were carried out: a) after 24 hours treatment of macrophages with NanoMBGs, cells were washed to eliminate the nanomaterial and then incubated 30 min with *Candida albicans* at the desired density (MOI 1 or MOI 5); b) simultaneous treatment of macrophages with NanoMBGs and *C. albicans* (MOI 1 or MOI 5) during 30 min. Control cells in the absence of nanomaterial or/and fungus were also analyzed.

**Table 2. Effects of FITC-NanoMBGs and *Candida albicans* on macrophage polarization towards M1 phenotype**

| Conditions | Percentage of CD80+ macrophages | |
|---|---|---|
| Control | 65 ± 3 | |
| *C. albicans* MOI 1 | 55 ± 2 *** (25% phagocytosis) | |
| *C. albicans* MOI 5 | 30 ± 2 *** (63% phagocytosis) | |
| | Percentage of CD80+ macrophages | |
| Conditions | a) Nano-MBGs 24 h | b) Nano-MBGs 30 min |
| Nano-MBGs | 56 ± 2 ** | 37 ± 2 *** |
| Nano-MBGs + *C. albicans* MOI 1 | 45 ± 2 *** (42%) | 29 ± 1 *** (47%) |
| Nano-MBGs + *C. albicans* MOI 5 | 26 ± 1 *** (59%) | 17 ± 1 *** (72%) |

As it can be observed in Table 2, 65% of control macrophages are CD80+. The presence of *Candida albicans* during 30 min induces a significant decrease (p<0.005) of this M1 population in a MOI dependent manner: 55% (MOI 1) and 30% (MOI 5). This effect is opposite to the observed effect of the fungus on the percentage of phagocytosis (in parentheses, in red) as the number of yeasts increases. The more pronounced effect is observed at MOI 5. At this condition, only a 30% of macrophages are CD80+, exhibiting also the highest percentage (63%) of phagocytosis.

When the two different treatments with NanoMBGs (a and b, previously indicated) were carried out, a significant decrease of $CD80^+$ population was also observed compared to the control macrophages (65%) and this effect was more pronounced at 30 min (37%, p<0.005) than at 24 hours (56%, p<0.01). These results indicate that these treatments with NanoMBGs induces on macrophages a decrease of their pro-inflammatory phenotype. Moreover, the addition of *Candida albicans* after the treatment with nanospheres, induces a more pronounced effect on CD80 marker expression in a MOI dependent manner. The a) treatment significantly reduces the M1 phenotype to 45% (p<0.005) and 26% (p<0.005) depending on the number of added yeasts (MOI 1 and 5, respectively). This effect was more pronounced after simultaneous addition (b treatment). In this case, the $CD80^+$ population significantly decreases to 29% (p<0.005) and 17% (p<0.005) at MOI 1 and MOI 5, respectively. These results show that the nanomaterial not only does not induce an inflammatory response but on the contrary, in contact with pathogens (like *C. albicans*), these nanospheres are able to modulate the M1 macrophage phenotype, decreasing this pro-inflammatory population. Taking into account these data and the relation with the macrophage phagocytosis capability observed after treatment with NanoMBGs (shown in red in the Table and described in Results 3.2), we can point out that the induction of this non inflammatory phenotype is related to a higher fungal phagocytosis. Recent studies on *C. albicans* phagocytosis by polarized M1 and M2 macrophages show that M1 macrophages have a lower phagocytic capability [48]. These results are also in agreement with previous studies with Raw 264.7 macrophages treated with these nanospheres, which induced also a decrease of the M1 phenotype [34]. On the other hand, we have observed a similar macrophage behavior after treatment with graphene oxide nanoparticles [8]. In previous studies, we have evaluated the effect of different GO nanosheets functionalized with poly(ethylene glycol)-amine and labelled with fluorescein isothiocyanate (FITC-PEG-GO) on macrophage and lymphocyte immune response. We observed that these GO nanosheets produced significant dose-dependent decreases of cell proliferation and IL-6 levels, revealing weak inflammatory properties of this nanomaterial, in the same way that we observed with nanospheres in the present work [4,8]. On the other hand, we have studied the polarization of murine peritoneal macrophages towards M1 and M2 phenotypes in the presence of these GO evidencing that FITC-PEG-GO uptake did not induce the macrophage polarization towards the M1 pro-inflammatory phenotype, promoting the control of the M1/M2 balance with a slight shift towards M2 reparative phenotype. Similar results have also been observed with the nanospheres in the present study. The uptake of NanoMBGs by macrophages produces a decrease of CD80$^+$ macrophage percentage and less IL-6 secretion, thus suggesting a non-inflammatory macrophage phenotype related to the simultaneous incorporation of NanoMBGs and the phagocytosis of the fungus [6,7]. Regarding the evaluation of the macrophage capacity to phagocytize the fungus *Candida albicans*, the effects of GO uptake on this macrophage activity has been evaluated in previous studies of our group. FITC-PEG-GO nanosheets were efficiently taken up by peritoneal macrophages inducing a significant increase of *C. albicans* phagocytosis by both pro-inflammatory macrophages (M1) and reparative macrophages (M2). On the other hand, after FITC-PEG-GO treatment and *C. albicans* infection, the percentages of GO$^+$ macrophages diminished when *Candida* uptake increased, thus suggesting the exocytosis of this nanomaterial. Similar results to those obtained in the present study indicates that the exocytosis of the nanomaterial is a dynamic mechanism favoring fungal phagocytosis [6,7].

Macrophages release different cytokines related with the modulation macrophage functions and the expression of cell surface markers [49]. In this sense, we have investigated the effects of the treatment with these nanospheres and *C. albicans* on the secretion of interleukin 6 (IL-6) by peritoneal macrophages, by measuring this pro-inflammatory cytokine in the culture medium.

We have selected the simultaneous treatment with NanoMBGs and *C. albicans* (MOI 1 and 5) for 30 min, as the most representative condition related to the observed CD80 decrease.

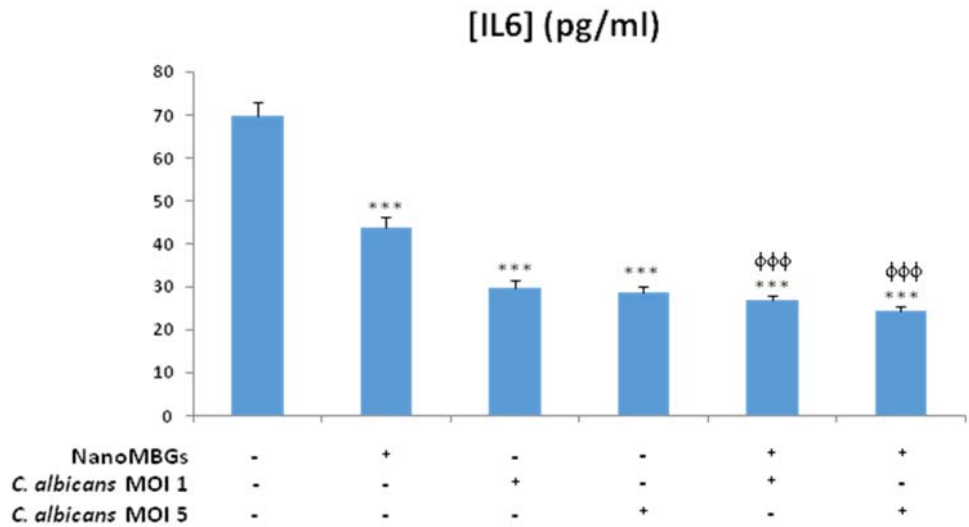

**Figure 6. Effects of NanoMBGs (50μg/mL) or/and *Candida albicans* at MOI 1 and MOI 5 on IL-6 production by peritoneal macrophages after 45 min of treatment.** Statistical significance: ***$p<0.005$ (comparison between each condition and control cells without nanomaterial and without fungus). $^{\phi\phi\phi}p<0.005$ (comparison between cells treated with nanomaterial + fungus and cells treated only with nanomaterial).

As it can be observed in Figure 6, a significant decrease of IL-6 was detected after treatment with NanoMBGs when compared with control macrophages ($p<0.005$). The infection of macrophages with the fungus also induces a significant decrease of the IL-6 levels ($p<0.005$). When comparing macrophages only treated with NanoMBGs and macrophages simultaneously treated with nanospheres and *Candida*, a similar and significant decrease ($p<0.005$) is detected. Again, the higher number of yeasts (MOI 5) in the presence of NanoMBGs produces the most pronounced IL-6 decrease, in agreement with the observed reduction of macrophage $CD80^+$ population (Table 2). Previous studies with graphene oxide nanosheets evidenced a significant dose-dependent decreases of cell proliferation and IL-6 secretion by primary splenocytes and Saos-2 osteoblasts, revealing weak inflammatory properties of this other nanomaterial [4,50].

The decrease of both CD80 expression and IL-6 secretion by macrophages due to the presence of NanoMBGs and *C. albicans*, evidences the absence of a macrophage pro-inflammatory phenotype thus suggesting the potential utility of these nanospheres for drug delivery without harmful side effects.

## 4. Conclusions

The present work highlights the importance of evaluating the effects of macrophage infection on their functional capability after the uptake of nanoparticles and *vice versa*. This study demonstrates that the exocytosis of NanoMBGs in the presence of C. albicans is a dynamic mechanism which favors the fungal phagocytosis, in agreement with previous studies with other nanoparticles [5,6]. In summary, the NanoMBG uptake by macrophages does not alters their phagocytosis capability and phenotypic plasticity. The following conclusions reflect the importance of studying the nanomaterial / macrophage / pathogen interface to design efficient therapies based on these nanocarriers potentially useful for drug delivery:

1. After assessing if the internalization of NanoMBGs decreases the phagocytosis capability of macrophages, we have demonstrated that the previous uptake of these nanospheres does not alter the macrophage function.
2. An efficient and maintained in time incorporation of the nanomaterial by macrophages was observed even in the presence of high doses of the fungus (MOI 5), thus ensuring the intracellular location of these nanospheres.
3. The significant decrease of CD80 expression and IL-6 secretion evidences a non-inflammatory macrophage phenotype related to the simultaneous incorporation of NanoMBGs and the fungus.

4. The detection by flow cytometry and confocal microscopy of both the nanomaterial and the fungus inside the macrophages, opens the way for future employment of these nanospheres for antifungal drug delivery. Moreover, the fact that the macrophage population with only the fungus inside is practically undetectable, reveals the high potential of this nanomaterial.

5. Once we had settled down a phagocytosis competition assay (between nanospheres and fungus), necessary to validate macrophage functional capability, our results have confirmed that macrophages clearly distinguish between the inert material and the live yeast, being a dynamic process and favoring the microbial phagocytosis.

The present study evidences that phagocytosis of macrophages can be closely connected to their functional state and, it has opened the way for future potential profits of these nanospheres for antifungal drug delivery.

## Acknowledgements

This study was supported by research grants from the Ministerio de Economía y Competitividad, Agencia Estatal de Investigación (AEI), Fondo Europeo de Desarrollo Regional (FEDER) (MAT2016-75611-R AEI/FEDER, UE) and 4129533-FEI-EU-2018 Project. MVR acknowledges funding from the European Research Council (Advanced Grant VERDI; ERC-2015-AdG Proposal No. 694160). The authors wish to thank the staff of the ICTS Centro Nacional de Microscopia Electrónica (Spain) and the Centro de Citometría y Microscopía de Fluorescencia of the Universidad Complutense de Madrid (Spain) for the assistance in the electron microscopy, flow cytometry and confocal microscopy studies.